\documentclass[11pt]{article}
\usepackage{dsfont, amssymb,amsmath,amscd,latexsym, amsthm, amsxtra,amsfonts, mathrsfs}
\usepackage{graphicx}
\usepackage[all]{xy}
\begin{document}
\baselineskip 18pt
\title{{ {An Improved FPRAS for Counting the Number of  Hamiltonian Cycles  in  Dense Digraphs}}}
\author{Jinshan \ Zhang\thanks{Electronic address:
zjs02@mails.tsinghua.edu.cn} \ and Fengshan Bai\thanks{Electronic
address: fbai@math.tsinghua.edu.cn}
\\ \small Department of Mathematical Sciences,
\  Tsinghua University,\\
\small Beijing 100084,  China }
\date{}
\maketitle

{\small\begin{center}\textbf{Abstract}\end{center}

We propose an improved algorithm for counting the number of
Hamiltonian cycles in a directed graph. The basic idea of the method
is sequential acceptance/rejection, which is successfully used in
approximating the number of perfect matchings in  dense bipartite
graphs. As a consequence, a new ratio of the number of Hamiltonian
cycles to the number of 1-factors is proposed. Based on this ratio,
we prove that our algorithm runs in expected time of $O(n^{8.5})$
for dense problems. This improves the Markov Chain Monte Carlo
method, the most powerful existing method, a factor of at least
$n^{4.5}(\log n)^{4}$ in running time. This class of dense problems
is shown to be nontrivial in counting, in the sense that they are
$\#$P-Complete.\\\\
\text{}}\\\\\\
\textbf{\large{ 1. Introduction}}

A Hamiltonian cycle is a closed directed path that visits each
vertex once and only once. In this paper we use digraph to denote
directed graph. Counting the number of Hamiltonian cycles is a very
challenging problem and has applications, for example, in quantum
physics \cite{CJ87}. Many intractable counting problems have been
added to the Valiant's\cite{Va79} list of $\#$P-Complete, which is a
natural correspondence  of the concept
 NP-Complete for decision problems. Efficient approximating schemes called fully
polynomial randomized approximation scheme(FPRAS) are naturally
considered for the hard problems in counting. If $M$ is the true
value, a randomized algorithm is called an FPRAS if it takes
polynomial time of size of inputs, $\varepsilon^{-1}$ and
$\log\delta^{-1}$ to obtain an output $\tilde{M}$. Here $\tilde{M}$
is the approximation of $M$, satisfying
\begin{displaymath}
P((1-\varepsilon)M\leq \tilde{M}\leq (1+\varepsilon)M)\geq 1-\delta.
\end{displaymath}

Due to the fact that the decision problem of whether a graph
contains a Hamiltonian cycle is NP-Complete, there would be no FPRAS
for counting the  Hamiltonian cycles for general graphs unless
NP=RP. Thus the FPRAS for  counting  Hamiltonian cycles are only
possible for special or restricted graphs, for example, elementary
recursive algorithms \cite{Ra94} for random digraphs; Markov Chain
Monte Carlo(MCMC) methods for dense $undirected$ graphs\cite{DFJ94},
for some random digraphs\cite{FS92}  and  random regular
graphs\cite{FJMRW96}.

Sequential acceptance/rejection method is introduced by Huber
\cite{Hu06} for counting the number of  the perfect matchings in a
dense regular bipartite graph. Recently the regularity
requirement is removed \cite{HL08}. The primary tool used in the
algorithm is the generalized Bregman's bound and the matrix scaling
method.

The MCMC algorithm presented for random digraphs in \cite{FS92} can
be naturally extended to dense digraphs. This algorithm is based on
sampling 1-factors of the digraphs and uses the self-reducing
method\cite{JVV86} to approximate the counting. Recently
Bez\'{a}kov\'{a} et. al.  present an algorithm that approximates the
number of 1-factors in $O(n^7(\log n)^4)$ expected time, via an
accelerating simulated annealing technique\cite{BSVV06}.

The ratio of the number of 1-factors to the number of Hamiltonian
cycles  is established to be $O(n^{1+1/(2\alpha-3/2)})$ in this
paper provided that the digraph is $\alpha n$ dense. Due to this
ratio and Bez\'{a}kov\'{a}'s results in \cite{BSVV06}, MCMC
method\cite{FS92} runs in an $O(n^{13}(\log n)^4)$ time when
$\alpha\geq.85$. Moreover, counting the number of Hamiltonian cycles
in such digraphs is shown to be still $\#$P-Complete.

Our algorithm for counting Hamiltonian cycles is built on the
acceptance/rejection algorithm in \cite{HL08} while a different
sequential sampling procedure is constructed to ensure that the
approximating target is the number of Hamiltonian cycles.

One of the remarkable advantages of acceptance/rejection method is
that it samples perfectly from a given set, which removes the
sampling error when the MCMC method is adopted. Hence, our algorithm
generates a weighted Hamiltonian cycle exactly according to its
weight from the set of Hamiltonian cycles of a weighted digraph. In
addition, this perfect sampling is only by-product when
acceptance/rejection is used to approximate counting, which means
the time used to sample a random Hamiltonian cycle can be used to
approximate the number of the Hamiltonian cycles without extra cost.
The main result of this paper is summarized  in the following.
\\\\
\textbf{Theorem M.}\emph{ For any $\varepsilon$, $\delta$ $\in$
$(0,1]$ and $\alpha\in (.75,1]$, there exists a randomized
approximation algorithm which provides an FPRAS for computing the
number of Hamiltonian cycles of $\alpha n$ dense digraphs. The same
algorithm approximates the number of Hamiltonian cycles by a factor
in $[1 - \varepsilon$, $1 + \varepsilon]$ with probability at least
$1 - \delta$ and has the complexity
$O(n^{2.5+.5/(2\alpha-1)+1/{(2\alpha-1.5)}}\varepsilon^{-2}\log(\delta^{-1}))$.
In particular, when $\alpha\geq .85$, the running time is bounded by
$O(n^{8.5})$.}\\

The remainder of the paper is structured as follows. In Section 2
some basic definitions, notations and lemmas are presented. In
Section 3 we describe the algorithm in details. Section 4
contributes to the complexity of the algorithm and the hardness of
counting. Further discussion and conclusion are proposed in Section
5.\\\\
\textbf{\large{2. Preliminaries}}


Consider a simple weighted digraph $G=(V, E)$ with the vertex set
$V=\{1,\cdots, n\}$ and the edge set $E$. Each edge $(i,j)\in E$ is
endowed with a positive weight $w_{ij}$. Let $|\ S \ |$  denote the
cardinality of any set $S$. The set of vertices pointing to $i$ is
denoted by $N^{-}(i,G)=\{j: (j,i)\in E\}$, and similarly that out of
$i$ by $N^{+}(i,G)=\{j: (i,j)\in E\}$. Indegrees and outdegrees of
the vertex $i$ are denoted by $\Delta^{-}(i)=|N^{-}(i,G)|$ ,
$\Delta^{+}(i)=|N^{+}(i,G)|$ respectively. Let
$\Delta(i)=\min(\Delta^{-}(i),\Delta^{+}(i))$ and $\Delta=\min_{i\in
V}\Delta(i)$. $G$ is called $\alpha n$ dense if $\Delta\geq\alpha n$
for an $\alpha>0$ given. Let $\oplus$ denote the symmetric
difference of two sets and $\lfloor n\rfloor$ denote the maximum
integer no more than $n$. $A/B$ is used to denote the set by
removing elements of $B$ from the set $A$. With a little abuse of
notation, $/$ also denote the quotient of two numbers. A Hamiltonian
cycle in $G$ is represented by

\begin{center}
$\mathcal {H}=(k_1,k_2,\cdots, k_n,k_1)$,
\end{center}

\noindent where $\{k_1,k_2,\cdots, k_n\}$ is a permutation of
$\{1,\cdots,n\}$ such that $(k_n,k_1)\in E$, and$(k_j,$ $k_{j+1})$
$\in E$, $j=1,\cdots, n-1$. The length of a cycle or path is defined
as the number of its edges that contains.

 An 1-factor is defined as a spanning directed subgraph of
G in which indegrees and outdegrees of each vertex are all one. An
example of an 1-factor is a spanning union of vertex disjoint
directed cycles. Obviously, a Hamiltonian cycle is a special
1-factor with only one cycle.  The weight $W(F)$ of an 1-factor $F$
with edge set $\{e\in E\}_{e\in F}$ is defined as $W(F)=\prod_{e\in
F}w_e$. The total weight $W(S)$ of the set $S$ of 1-factors are
defined as $W(S)=\sum_{F\in S}W(F)$. Let $W_F(G)$ and $W_H(G)$
denote the total weight of all the 1-factors and Hamiltonian cycles
in $G$ respectively. It is easy to see if $w_{ij}=1$ for all
$(i,j)\in E$, then $W_F(G)$ and $W_H(G)$ are the number of 1-factors
and Hamiltonian cycles in $G$ respectively.

Let $A_G$ be the adjacent matrix associated with $G$ where
$A_G(i,j)=w_{ij}$ if $(i,j)\in E$ and $A_G(i,j)=0$ otherwise. For an
$n\times n$ matrix $A$, where $n$ is the order of $A$, we use
notation $A_{ij}$ to denote the $(n-1)\times (n-1)$ matrix obtained
from $A$ by removing  row $i$ and column $j$. If there is no
confusion, $A^{'}_{ij}$ or $(A_{ij})^{'}$ denotes the $(n-1)\times
(n-1)$ matrix obtained from $A$ by, first permutating  row $i$ and
row $j$ and then removing row $j$ and column $j$. Next we will
define two quantities on the matrix $A_G$ which are related to
1-factors
and Hamiltonian cycles respectively.\\\\
\textbf{Definition 1.} The permanent of an $n\times n$ matrix
$A=(A(i,j))_{n\times n}$ is
\begin{displaymath}
\operatorname{per}(A)=
\sum\limits_{\sigma}\prod\limits_{i=1}^{n}A(i,\sigma(i)),
\end{displaymath}
where $\sigma$ ranges over all the permutations of
$\{1,\cdots,n\}$.\\\\
\textbf{Definition 2.} The Hamilton of an $n\times n$ matrix
$A=(A(i,j))_{n\times n}$ is defined as
\begin{displaymath}
\operatorname{ham}(A)= \sum\limits_{\{k_1 \cdots,
k_{n-1}\}}A(k_1,1)A(k_2,k_1)\cdots A(k_{n-1},k_{n-2})A(1,k_{n-1}),
\end{displaymath}
where $\{k_1 \cdots, k_{n-1}\}$ ranges over all the permutations of
$\{2,\cdots,n\}$ when $n\geq 2$, and $\operatorname{ham}(A)=A(1,1)$, if $n=1$.\\

By the definition of permanent and Hamilton, it is not difficult to
see that  $\operatorname{per}(A)\geq \operatorname{ham}(A)$ if the
entries of $A$ are all nonnegative. Suppose $A=A_G$. For any
permutation $(k_1 \cdots, k_{n-1})$ of $(2,\cdots,n)$,
$A_G(1,k_{n-1})$, $A_G(k_{n-1},k_{n-2})$, $\cdots$, $A_G(k_2,k_1)$,
$A_G(k_1,1)$ are the edge weight of the Hamiltonian cycle
$(1,k_{n-1},\cdots, k_1,1)$ in $G$ if and only if they are all
positive. Therefore, we have
\begin{displaymath}
W_H(G)=\operatorname{ham}(A_G).
\end{displaymath}
Note that the diagonal entries of $A_G$ are all zero, and for any
permutation $\sigma$ over $\{1,2,\cdots,n\}$, $A(i,\sigma(i))> 0$,
$i=1,\cdots,n$ if and only if their corresponding edges in $G$ form
an 1-factor of $G$. Hence
\begin{displaymath}
W_F(G)=\operatorname{per}(A_G).
\end{displaymath}

Next we present the Laplacian  expansion formulas for the permanent
and the Hamilton.\\\\
\textbf{Lemma 3.} \emph{Let $A=(A(i,j))_{n\times n}$ be an $n\times
n$ matrix. The permanent of empty matrix is set to be 1. Then
\begin{displaymath}
\operatorname{per}(A)=\sum\limits_{i=1}^{n}A(i,1)\operatorname{per}(A_{i1}).
\end{displaymath}}\\\\
\textbf{Lemma 4.} \emph{Let $A=(A(i,j))_{n\times n}$ be an $n\times
n$ matrix, $n\geq 2$. Then
\begin{displaymath}
\operatorname{ham}(A)=\sum\limits_{i=2}^{n}A(i,1)\operatorname{ham}(A^{'}_{i1}).
\end{displaymath}}

For the permanent, this expansion is well known. For the Hamilton,
the formula is very similar and \cite{Ra94} proposes a combinatorial
proof when each edge weight of the digraph is one. Regarding its
importance in our algorithm, a proof in terms of matrix is presented
below. We emphasize Lemma 4 is crucial in the sequential sampling
procedure which is different from the one used in \cite{HL08}, and
ensures our
algorithm to approximate the number of Hamiltonian cycles.\\\\
\textbf{Proof of Lemma 4.} We proceed to prove the lemma by induction on $n$, the order of the matrix.\\
The case $k=2$ is trivial.\\
Suppose Lemma 4 holds for $k=n-1$. \\
Consider $k=n$. Since
\begin{displaymath}
\operatorname{ham}(A)=\sum\limits_{i=2}^{n}A(i,1)\sum\limits_{\{k_2
\cdots, k_{n-1}\}}A(k_2,i)A(k_3,k_2)\cdots
A(k_{n-1},k_{n-2})A(1,k_{n-1}),
\end{displaymath}
it is sufficient to show that
\begin{displaymath}
\operatorname{ham}(A^{'}_{i1})=\sum\limits_{\{k_2 \cdots,
k_{n-1}\}}A(k_2,i)\cdots A(k_{n-1},k_{n-2})A(1,k_{n-1}),
\end{displaymath}
for $i=2,\cdots,n$, where $\{k_2 \cdots, k_{n-1}\}$ goes over all
the permutations of $\{2 \cdots, n\}/i$. Considering the definition
of $A^{'}_{i1}$, the  row $i-1$ of $A^{'}_{i1}$ is the first row of
$A$ except removing the first element, and

\hskip 20mm $A^{'}_{i1}(k_2-1,i-1)=A(k_2,i)$,
$\cdots$ $\cdots$,

\hskip 20mm $A^{'}_{i1}(k_{n-1}-1,k_{n-2}-1)=A(k_{n-1},k_{n-2})$,
and

\hskip 20mm $A^{'}_{i1}(i-1,k_{n-1}-1)=A(1,k_{n-1})$.

\noindent By the hypothesis of the induction, the order of
$A^{'}_{i1}$ is $n-1$, then
\begin{displaymath}
\begin{split}
\operatorname{ham}(A^{'}_{i1})&=\sum\limits_{\{k^{'}_2 \cdots,
k^{'}_{n-1}\}}A^{'}_{i1}(k^{'}_2,i-1)\cdots
A^{'}_{i1}(k^{'}_{n-1},k^{'}_{n-2})A^{'}_{i1}(i-1,k^{'}_{n-1})\\
&=\sum\limits_{\{k_2 \cdots,
k_{n-1}\}}A^{'}_{i1}(k_2-1,i-1)A^{'}_{i1}(k_{n-1}-1,k_{n-2}-1)A^{'}_{i1}(i-1,k_{n-1}-1)\\
&=\sum\limits_{\{k_2 \cdots, k_{n-1}\}}A(k_2,i)\cdots
A(k_{n-1},k_{n-2})A(1,k_{n-1}),
\end{split}
\end{displaymath}
where $\{k^{'}_2 \cdots, k^{'}_{n-1}\}$ and $\{k_2 \cdots,
k_{n-1}\}$ go over all the permutations of $\{1,\cdots,
n-1\}/\{i-1\}$
and  $\{2,\cdots, n\}/\{i\}$ respectively. This completes the proof of Lemma 4. \ \ \ $\Box$\\\\
\textbf{Hamiltonian Recovery} Let $A=(A(i,j))_{n\times n}$ be an
$n\times n$ positive matrix.
The following procedure is applied to selecting elements from $A$ (The first two steps are given explicitly).
We call this procedure Selecting Hamiltonian Cycle(SHC for simplicity). \\\\
\textbf{Step 1.} Let $A^{1}=A$. Choose a natural number $1<j_1\leq
n$, denote $\pi(1)=j_1$ and select $A^1(\pi(1),1)$. \\
\textbf{Step 2.} Let $A^{2}=(A^{1}_{j_11})^{'}$. Choose a natural
number $1<j_2\leq n-1$, denote $\pi(2)=j_2$ and select $A^2(\pi(2),1)$.\\
Similarly  $A^{k}$, $\pi(k)$ and $A^k(\pi(k),1)$, $1\leq k\leq n-1$
can be obtained in Step k iteratively. Since $A^{n}$ has only one
entry, let $A^{n}=(A^{n-1}_{j_{n-1}1})^{'}$
, $\pi(n)=j_n=1$ and select $A^n(\pi(n),1)$.\\

By Lemma 4, the set of selected elements $A^k(\pi(k),1)$,
$k=1,\cdots,n$, from the above procedure forms the edge weight of a
Hamiltonian cycle in $G$ if $A=A_G$. If
$\pi(1),\pi(2),\cdots,\pi(n)$ is given, we provide a simple
algorithm to determine which Hamiltonian cycle in $G$ is selected.
This process is called Hamiltonian Recovery.

The input of the algorithm is $\pi=(\pi(1),\pi(2),\cdots, \pi(n))$.
We illustrate how to recover an entry in $A^{2}$ if $\pi(1)$ is
given. Let $A^{2}(i,j)$ be any entry in $A^{2}$. Since
$A^{2}=(A^{1}_{\pi(1)1})^{'}$ and recall the definition of
$(A^{1}_{\pi(1)1})^{'}$, which is obtained by, first permutating row
$\pi(1)$ and the first row and then removing the first row and first
column. Hence, if $i=\pi(1)-1$ then $(1, j+1)$ is the position where
$A^{2}(i,j)$ lies of $A^{1}$; otherwise $(i+1,j+1)$ is the position
where $A^{2}(i,j)$ lies of $A^{1}$. Hence, if the vector
$(\pi(1),\cdots,\pi(k-1))$ is given from the SHC procedure, the
position of $A^{k}(\pi(k),1)$ in $A^1$ can be found recursively by
determining its position in $A^{k-1}$, then in $A^{k-2}$, and
finally in $A^1$. Since at each step of the SHC procedure an element
is selected from the first column, $A^{k}(\pi(k),1)$ must lie in
column $k$ of $A^1$.

If $(1,k_1,\cdots,k_{n-1},1)$ is the corresponding Hamiltonian cycle
of $\pi=(\pi(1)$, $\pi(2)$, $\cdots,\pi(n))$,  then $k_i$ can be
obtained from $k_{i+1}$ since the element $A(k_i,k_{i+1})$ is
selected in Step $k_{i+1}$ of the SHC procedure, or equivalently
$(k_i,k_{i+1})$ is the position of $A^{k_{i+1}}(\pi(k_{i+1}),1)$ in
$A$, $i=1,2,\cdots,n-2$. Obviously, $k_{n-1}=\pi(1)$. By this simple
procedure, it takes $O(n^2)$ time to recover all the positions of
$A^{k}(\pi(k),1)$, $1\leq
k\leq n$. We present the recovery algorithm explicitly.\\\\
\textbf{Hamiltonian Recovery Algorithm}\\
$Input:$ The vector $(\pi(1),\pi(2),\cdots,\pi(n))$. \\
$Output:$ A Hamiltonian cycle $(1,k_1,\cdots,k_{n-1},1)$.\\
\textbf{Step 1:} Set $k_{n-1}=\pi(1)$;

\ \ \ \ \ \ \ \  \textbf{For} $i=n-2$ to $1$

\ \ \ \ \ \ \ \ \ \ \ \ \ \ \ \ \ \ Set $a=\pi(k_{i+1})$;

\ \ \ \ \ \ \ \ \ \ \ \ \ \ \ \ \ \ \textbf{For} $j=k_{i+1}$ to $2$

\ \ \ \ \ \ \ \ \ \ \ \ \ \ \ \ \ \ \ \ \ \ \ \ \textbf{If}
$a=\pi(j-1)-1$; Set $a=1$;

\ \ \ \ \ \ \ \ \ \ \ \ \ \ \ \ \ \ \ \ \ \ \ \ \textbf{Else} Set
$a=a+1$;

\ \ \ \ \ \ \ \ \ \ \ \ \ \ \ \ \ \ \ \textbf{End};

\ \ \ \ \ \ \ \ \ \ \ \ \ \ \ \ \ \ Set $k_{i}=a$;

\ \ \ \ \ \ \ \  \textbf{End};

\ \ \ \ \ \ \ \  Goto Step 2;\\
\textbf{Step 2:} Output $(1,k_1,\cdots,k_{n-1},1)$.\\

For simplicity,  let $HR(\pi)$ denote the output of the Hamiltonian
Recovery Algorithm when the input is
$\pi=(\pi(1),\pi(2),\cdots,\pi(n))$.\\\\
\textbf{\large{3. Algorithms for Counting}}

One main tool in our algorithm is a generalized version of Bregman's
bound for the permanent below, which generalized an inequality of
Soul \cite{So00} and proved in \cite{HL08}. For more application of
other generalization of Bregman's bound for designing new algorithms
or improving efficiency of algorithms, we refer to
\cite{LSW00,Sa06}. Let
\begin{displaymath}
g(r)=\left\{ \begin{aligned}
 r+(1/2)\log r+e-1,\ \ r\geq 1 \\
 1+(e-1)r,\ \ \ \ \ \ \ \ \ \ \ r\in[0,1].
\end{aligned} \right.
\end{displaymath}\\\\
\textbf{Lemma 5.} (\cite{HL08}) \emph{Let $A$ be an $n\times n$
matrix with entries in $[0,1]$. Denote $r(i)$ the sum of  row $i$ of
$A$. Define $Br(A)=\prod\limits_{i=1}^{n}(g(r(i))/e)$, then
\begin{displaymath}
Br(A)\geq \sum\limits_{i=1}^{n}A(i,1)Br(A_{i1}).
\end{displaymath}
In particular, by Lemma 3, $\operatorname{per}(A)\leq Br(A)$.}\\\\

Chernoff's bound is useful in our algorithm, and one form of
that is given bellow\cite{MR95}.\\\\
\textbf{Lemma 6.} \emph{Let $x_1, x_2,\cdots, x_t$ be identical
independent distributed(i.i.d.) Bernoulli random variables with
$P(x_1=1)=p$ and $P(x_1=0)=1-p$, $p>0$, then for any
$0\leq\varepsilon\leq 2e-1$,
\begin{displaymath}
P(|\sum\limits_{i=1}^t x_i -tp|>\varepsilon tp)\leq
e^{-tp\varepsilon^{2}/4}.
\end{displaymath}}

For simplicity, in this section we only consider the digraph $G$
with all edge weight equalling one. Hence the adjacent matrix $A_G$
is a 0-1 matrix and $\operatorname{ham}(A_G)$ is the number of
Hamiltonian cycles in $G$.  $G$ is also restricted to be $\alpha n$
dense, $\alpha\geq.75$. It is known \cite{BM76} that if $G$ is $.5n$
dense, $G$ must contain a Hamiltonian cycle and the proof can be
easily modified to give an $O(n^2)$ algorithm to construct a
Hamiltonian cycle. Hence $\operatorname{ham}(A_G)\geq 1$. By the
definition of Hamilton, if we change any zeros in $A_G$ to
$\gamma=(\varepsilon/3)((n-1)!)^{-1}$, $\operatorname{ham}(A_G)$
increases by at most a factor of $1+ \varepsilon/3$.

Now we introduce the basic idea of acceptance/rejection method for
the counting problem. Suppose $S$ is a large set and each element in
it with positive weight. The target is to approximate the total
weight of all the elements in $S$. First select a suitable large $M$
such that $M> \sum_{b\in S}w(b)$. The main idea of
acceptance/rejection method for approximation is to design a
procedure to sample a random element $x$ from the set $S$ with the
successful probability $P(x=a)=\frac{w(a)}{M}$, where $w(a)$ is the
weight of $a\in S$, and failing probability $P(x\notin
S)=1-\frac{\sum_{b\in S}w(b)}{M}$. At each time, if a random element
$a$ is successfully selected from $S$, we say acceptance or $a$ is
accepted, and if no element is selected from $S$, we say rejection.
Hence, at each time the probability of acceptance is
$\frac{\sum_{b\in S}w(b)}{M}$ and probability of rejection
$1-\frac{\sum_{b\in S}w(b)}{M}$. With some fundamental statistical
knowledge, the total weight of $S$ can be approximated by multipling
$M$ and the ratio of acceptance over all the samplings. For our
purpose, generalized Bregman's bound in Lemma 5 provides such a
suitable large $M$,  and self-reducing method for counting
Hamiltonian cycles naturally proposes such a sampling procedure,
which is sequential sampling procedure guaranteed by Lemma 4. For
more details about sequential acceptance/rejection method, we refer
to \cite{HL08}.

In order to make use of the generalized Bregman's bound in Lemma 5,
before resuming the acceptance/rejection algorithm, we need to scale
the matrix $A_G$ to nearly be doubly stochastic and make each entry
in [0,1]\cite{KK96}. Hence the algorithm has two
phases. \\\\
\textbf{Sub Algorithm I. Scale Matrix}\\
$Input:$ $A_G$, $\varepsilon$ \\
$Output:$ $X$, $Y$, $Z$, $C$ \\
\textbf{Step 1:} Set $A_G(i,j)= (\varepsilon/3)((n-1)!)^{-1}$ if $A_G(i,j)=0$ for all $i$, $j$, goto Step 2;\\
\textbf{Step 2:} Using matrix scaling to find diagonal matrix $X$,
$Y$ such that the row and
column sums of $B=XA_GY$ in ($1-(.1)n^{-2},1+(.1)n^{-2})$, goto Step 3; \\
\textbf{Step 3:} Let $Z$ be a diagonal matrix with
$Z(i,i)=\min_jB(i,j)^{-1}$ for
$i=1,\cdots n$, goto Step 4;\\
\textbf{Step 4:} $C=ZB$.\\

After matrix scaling, matrix $C$ satisfies the requirement of
generalized Bregman's bound. Sequential acceptance/rejection method
can be used to estimate $\operatorname{ham}(C)$.
Note that the matrix $C$ is corresponding to a weighted digraph denoted by $G_C$.\\\\
\textbf{Sub Algorithm II.  Approximating Hamilton via Acceptance/Rejection}\\
$Input:$ $X$, $Y$, $Z$, $C$, $\varepsilon$, $\delta$ $N$.\\
$Output:$  $\mathcal {H}_1,\cdots,\mathcal {H}_s$;
$\widetilde{\operatorname{ham}}(A_G)$
 the estimator of $\operatorname{ham}(A_G)$. \\
\textbf{Step 5:} Set $t=4N(\varepsilon/2)^{-2}\log(\delta^{-1})$,
$l=\prod\limits_{i=1}^n(X(i,i)Y(i,i)Z(i,i))$, $D=C$, $k=0$ and
$s=0$, goto
Step 6;\\
\textbf{Step 6:} Set $r=$ order of $D$;

\ \ \ \ \ \ \ \textbf{If} $r=1$;

\ \ \ \ \ \ \ \ \ Set $p(1)= D/Br(D)$ and $p(0)=1-p(1)$;

\ \ \ \ \ \ \ \ \ Choose $I$ from $\{0,1\}$ according to
                         $P(I=i)=p(i)$, $i=0, 1$;

\ \ \ \ \ \ \ \ \ \ \ \ \ \ \ \ \textbf{If} $I>0$; Set $\pi(n)=1$,
$s=s+1$, $k=k+1$
                             and $\mathcal{H}_s=HR(\pi)$;

 \ \ \ \ \ \ \ \ \ \ \ \ \ \ \ \ \ \ \ \ \ \ \textbf{If} $k<t$; Set $D=C$, goto Step 6; Otherwise
 goto Step 7;

\ \ \ \ \ \ \ \ \ \ \ \ \ \ \ \  \textbf{Else} $I=0$; Set $k=k+1$;

 \ \ \ \ \ \ \ \ \ \ \ \ \ \ \ \ \ \ \ \ \ \ \textbf{If} $k<t$; Set $D=C$, goto Step 6;
Otherwise goto Step 7;

\ \ \ \ \ \ \  \textbf{Else} $r>1$;

\ \ \ \ \ \ \ \ \ \  Set $p(i)= D(i,1)Br(D^{'}_{i1})/Br(D)$ for
$i=2,\cdots,r$ and $p(0)=1-\sum\limits_{i=2}^{r}p(i)$;

\ \ \ \ \ \ \ \ \ \ \ Choose $I$ from $\{0, 2, 3,\cdots, r\}$
according to $P(I=i)=p(i)$, $i=0, 2,\cdots, r$;

\ \ \ \ \ \ \ \ \ \ \ \ \ \ \ \  \textbf{If} $I>0$; Set
$\pi(n+1-r)=I$
                                and $D=D^{'}_{I1}$, goto Step 6;

\ \ \ \ \ \ \ \ \ \ \ \ \ \ \ \  \textbf{Else} $I=0$; Set $k=k+1$

 \ \ \ \ \ \ \ \ \ \ \ \ \ \ \ \ \ \ \ \ \ \ \textbf{If} $k<t$; Set $D=C$, goto Step 6; Otherwise goto
                              Step 7;\\
\textbf{Step 7:}
$\widetilde{\operatorname{ham}}(A_G)=l^{-1}st^{-1}Br(C)$.\\

The procedure of sampling elements in Step 6 is the same as SHC
procedure except selecting an element with certain probability or
rejection when $I=0$ is selected. The output
$\mathcal{H}_i$, $1\leq i \leq s$, is accepted by the algorithm.\\\\
\textbf{Theorem 7.} \emph{Let $\mathcal {H}_1,\cdots,\mathcal {H}_s$
and $\widetilde{\operatorname{ham}}(A_G)$ be the output of Sub
Algorithm II. If we set $N=Br(C)/\operatorname{ham}(C)$ in the same
algorithm, and let $H$ be a random variable recovered from a random
$\pi$ of Sub Algorithms II and $S$ denote the set of all the
possible accepted hamiltonian cycles, then
\begin{displaymath}
P(H=\mathcal {H}_1|H\in S)=W(\mathcal {H}_1)/W_H(G_C)
\end{displaymath}
and
\begin{displaymath}
P((1-\varepsilon)\operatorname{ham}(A_G)\leq
\widetilde{\operatorname{ham}}(A_G)\leq
(1+\varepsilon)\operatorname{ham}(A_G))\geq 1-\delta.
\end{displaymath}}\\
\textbf{Proof.} First, we check $p(0)\geq 0$ at each level of  Step
6, which guarantees the proceeding of the algorithm.  By the
definition of $D^{'}_{i1}$ and $D_{i1}$, obviously,
$Br(D^{'}_{i1})=Br(D_{i1})$. Using Lemma 5, it is easy to see
\begin{displaymath}
\begin{split}
\sum\limits_{i=2}^nD(i,1)Br(D^{'}_{i1})&=\sum\limits_{i=2}^nD(i,1)Br(D_{i1})\\
&\leq \sum\limits_{i=1}^nD(i,1)Br(D_{i1})\\
&\leq Br(D).
\end{split}
\end{displaymath}
Hence $p(0)\geq 0$. Suppose $\mathcal{H}_1=HR(j)$,
$j=(j_1,\cdots,j_n)$. Following the path in which $\mathcal {H}_1$
is selected, and using the notation in SHC procedure, then
$C^{i+1}=(C^{i}_{j_i1})^{'}$, $i=1,\cdots,n-1$ and $C^1=C$, we
have\\
\begin{displaymath}
 P(\pi(k)=j_k)=
\frac{C^k(j_k,1)Br((C^k_{j_k1})^{'})}{Br(C^k)},
\end{displaymath}
 where $k=1,2,\cdots,n-1$ and $P(\pi(n)=j_n)=\frac{C^n(j_n,1)}{Br(C^n)}$. \\
Since the selection at each level in Step 6 is independent of the
other, the probability of selecting $\mathcal {H}_1$ is the
telescoping product. Noting that
$C^{i+1}=(C^{i}_{j_i1})^{'}$, $i=1,\cdots,n-1$ and $C^1=C$, then \\
\begin{displaymath}
\begin{split}
P(H=\mathcal{H}_1)&=P(\pi=j)=\prod\limits^n_{k=1}P(\pi(k)=j_k)
=\frac{\prod_{i=1}^{n}C^i(j_i,1)}{Br(C)} =\frac{W(\mathcal
{H}_1)}{Br(C)}.
\end{split}
\end{displaymath}

Since each Hamiltonian cycle in $G_C$ can be accepted with certain
probability proportional to its weight, the acceptance set $S$ is
the set of all the Hamiltonian cycles in $G_C$. Then
\begin{displaymath}
P(H\in S)=\sum\limits_{\mathcal{H}\in
G_C}P(H=\mathcal{H})=\frac{W_H(G_C)}{Br(C)}.
\end{displaymath}
Hence,
\begin{displaymath}
P(H=\mathcal{H}_1|H\in S)=\frac{W(\mathcal{H}_1)}{W_H(G_C)}.
\end{displaymath}

In Sub Algorithm II, let $x_k$, $1\leq k\leq t$, denote the
indication function of acceptance or rejection in Step 6, that is,
$x_k=1$ if a Hamiltonian cycle is accepted and $x_k=0$ otherwise.
Obviously, $x_k$, $1\leq k\leq t$, are i.i.d. Bernoulli random
variables with
$P(x_1=1)=p=W_H(G_C)/Br(C)=\operatorname{ham}(C)/Br(C)$. Let
$A^{\varepsilon}_G$ be the matrix obtained in Step 1 of Sub
Algorithm I. Hence, by Lemma 6 and noting
$t=4N(\varepsilon/2)^{-2}\log(\delta^{-1})$, where
$N=Br(C)/\operatorname{ham}(C)$, a simple calculation shows
\begin{displaymath}
P((1-\varepsilon/2)\operatorname{ham}(A^{\varepsilon}_G)\leq
\widetilde{\operatorname{ham}}(A_G)\leq
(1+\varepsilon/2)\operatorname{ham}(A^{\varepsilon}_G))\geq
1-\delta.
\end{displaymath}
Noting $\operatorname{ham}(A_G)\leq
\operatorname{ham}(A^{\varepsilon}_G)\leq
(1+\varepsilon/3)\operatorname{ham}(A_G)$, thus the proof completes. \ \ $\Box$\\\\\\
\textbf{\large{4. Complexity and Hardness of Counting}}\\\\
\textsl{4.1. Complexity of the Algorithm}

Due to ellipsoid method\cite{KK96}, the running time of matrix
scaling is $O(n^4\log n)$. So the complexity of  Sub Algorithm I is
$O(n^4\log n)$.

The time of repeating Step 6 in Sub Algorithm II is $t$ $=$
$O(Br(C)/\operatorname{ham}(C))$, and for each time the running time
is $O(n^2)$, hence, the complexity of Sub Algorithm II is
$O(n^2*t)=O(n^2Br(C)/\operatorname{ham}(C))$, where
$\varepsilon^{-2}\log\delta^{-1}$ has been put into the term
$O(\cdot)$ for simplicity. As we know, the Hamiltonian Recovery
Algorithm takes $O(n^2)$ time. After removing Hamiltonian Recovery
procedure, the total running time of Sub Algorithm II is still
$O(n^2*t)$, thus if approximating $\operatorname{ham}(A_G)$ is only
the purpose, outputting the Hamiltonian cycle is the byproduct of
Sub Algorithm II.

 If the digraph $G$ is $\alpha n$
dense, $\alpha
>.5$, an important result given by Huber\cite{HL08} is
\begin{displaymath}
Br(C)/\operatorname{per}(C)=O(n^{-.5+.5/(2\alpha-1)}).
\end{displaymath}
Note that\\
\begin{displaymath}
\begin{split}
 \frac{\operatorname{per}(C)}{\operatorname{ham}(C)}&=
  \frac{\prod\limits_{i=1}^n(X(i,i)Y(i,i)Z(i,i))\operatorname{per}(A^{\varepsilon}_G)}
 {\prod\limits_{i=1}^n(X(i,i)Y(i,i)Z(i,i))\operatorname{ham}(A^{\varepsilon}_G)}
 =\frac{\operatorname{per}(A^{\varepsilon}_G)}{\operatorname{ham}(A^{\varepsilon}_G)}.
\end{split}
\end{displaymath}
If the digraph $G$ is at least $.5n$ dense, then changing any zeros
in $A_G$ to $\varepsilon n^{-3}$ increases $\operatorname{per}(A_G)$
by at most a factor of $1+\varepsilon$ \cite{JS89}. Then
\begin{displaymath}
\begin{split}
  \frac{\operatorname{per}(C)}{\operatorname{ham}(C)}
  =\frac{\operatorname{per}(A^{\varepsilon}_G)}{\operatorname{ham}(A^{\varepsilon}_G)}
  \leq\frac{(1+\varepsilon/3)\operatorname{per}(A_G)}{\operatorname{ham}(A_G)}
 =O(\frac{\operatorname{per}(A_G)}{\operatorname{ham}(A_G)}).
\end{split}
\end{displaymath}
Hence, the total running time of our algorithm is
\begin{equation}
\begin{split}
O(n^4\log n+ n^2\frac{Br(C)}{\operatorname{ham}(C)})&= O(n^4\log n+
n^2\frac{Br(C)}{\operatorname{per}(C)}\frac{\operatorname{per}(C)}{\operatorname{ham}(C)})\\
&=O(n^4\log n+
n^{1.5+.5/(2\alpha-1)}\frac{\operatorname{per}(A_G)}{\operatorname{ham}(A_G)}).
\end{split}
\end{equation}
Now we present combinatorial argument on the bound of
$\frac{\operatorname{per}(A_G)}{\operatorname{ham}(A_G)}$(Recall
$A_G$ is a 0-1 matrix and all the edge weight of $G$ equals one).
The methodology
is analogous to the approach for undirected graphs given by Dyer et.al.\cite{DFJ94}. \\\\
\textbf{Lemma 8.} (\cite{DFJ94}) \emph{Let $n$ be a natural number
and $\beta$ a positive number. Let $k_0 = \max(\lfloor\beta\log
n\rfloor,1)$ and
$g(k)=n^{\beta}k!(\beta\log n)^{-k}$, define\\
\begin{displaymath}
f(k)=\left\{ \begin{aligned}
 g(k),\ \ \ \ k\leq k_0 \\
 g(k_0),\ \ k>k_0.
\end{aligned} \right.
\end{displaymath}
Then $f(k-1)\geq (\beta\log n)k^{-1}f(k)$;  and $f(k)\geq 1$ for any
$k$.}\\\\
\textbf{Proof.} If $k\leq k_0$, $f(k-1)=g(k-1)=(\beta\log
n)k^{-1}g(k)=(\beta\log n)k^{-1}f(k)$;

If $k> k_0$, then $\beta\log n/k\leq 1$. Hence
$$f(k-1)=g(k_0)\geq(\beta\log n)k^{-1}g(k_0)=(\beta\log
n)k^{-1}f(k).$$
Thus $f(k)\geq f(k_0)$, we have\\
\begin{displaymath}
 \frac{1}{f(k)}\leq \frac{1}{f(k_0)}\leq\frac{(\beta \log
n)^{k_0}}{n^{\beta}(k_0)!}\leq
n^{-\beta}\sum\limits_{k=0}^{\infty}\frac{(\beta \log
n)^{k}}{(k)!}\leq n^{-\beta}e^{\beta \log n}=1.\ \ \ \Box
\end{displaymath}\\\\
\textbf{Theorem 9.} \emph{Suppose $\alpha\in(.75,1]$. Let $G$ be an
$\alpha n$ dense digraph and $F_k$ the set of 1-factors in $G$
containing exactly $k$ cycles, $1\leq k\leq \lfloor n/2\rfloor$.
Note that $F_1$ is the set of Hamiltonian cycles in $G$. Let
$F=\bigcup_kF_k$. Then
\begin{displaymath}
\frac{|F|}{|F_1|}=O(n^{1+1/(2\alpha-1.5)}).
\end{displaymath}}

With this theorem, we prove the main result of this paper Theorem M.\\\\
\textbf{Proof of Theorem M.} By theorem 9, since
$|F|/|F_1|=\operatorname{per}(A_G)/\operatorname{ham}(A_G)$, and
noting (1), therefore Theorem M follows immediately.\\

\begin{figure}[ht]
\centering
\includegraphics[scale=.75]{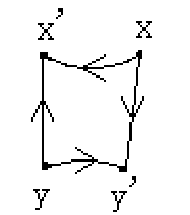}
\caption{\small{The symmetric differences $\overline{C_4}=E\oplus
E^{'}$}}
\end{figure}

\begin{figure}[ht]
\centering
\includegraphics[scale=.75]{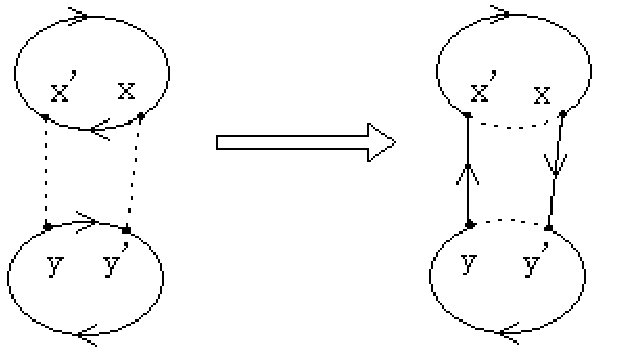}
\caption{\small{$E^{'}$ is obtained by coalescing two cycles of $E$
into a single cycle}}
\end{figure}
Now we proceed to prove Theorem 9.\\\\
\textbf{Proof of Theorem 9.} We construct a new weighted digraph
$\Psi=(F,K)$. $K$ is defined as follows.
\begin{displaymath}
K=\{(E,E^{'}): E\in F_k, E^{'}\in F_{k^{'}}, k^{'}<k \ \
\operatorname{and} \ \ E\oplus E^{'}\cong \overline{C_4} \},
\end{displaymath}
where $\overline{C_4}$ is a graph with four vertices and four edges,
in which two vertices have indegrees two, outdegrees zero, and the
other two vertices have indegrees zero, outdegrees two(See Figure
1). The four edges belong to $E$ and $E^{'}$ alternatively. To avoid
the confusion with vertices and edges in $G$, we call the $nodes$
and $arcs$ corresponding to $F$ and $K$ in $\Psi$. Observe also that
if $(E,E^{'})\in K$ is an arc of $\Psi$, $E^{'}$ can be obtained
from $E$ by deleting two edges and adding two others, and that this
operation can decrease the number of cycles by one(See Figure 2).
Hence every arc $(E, E^{'})$ is directed from a node $E$ in some
$F_k$ to a node $E^{'}$ in $F_{k-1}$.

The proof strategy is to define a positive weight function $w$ on
the arcs set $K$ such that the total weight of arcs leaving each
node $E\in F/F_{1}$ is at least one greater than the total weight of
arcs entering $E$. Denote $w^{+}(S)$ and $w^{-}(S)$  the total
weight leaving and entering a node set $S$ in $\Psi$ respectively,
the strategy ensures
\begin{displaymath}
w^{-}(F_k) + |F_k|=\sum_{E\in F_k}(w^{-}(E)+1)\leq \sum_{E\in
F_k}w^{+}(E)=w^{+}(F_k),\ \ \ \ k\geq 2.
\end{displaymath}
Hence,
\begin{displaymath}
 w^{-}(F_1)=w^{+}(F_2)=\sum_{k\geq2}(w^{+}(F_k)-w^{-}(F_k))\geq
\sum_{k\geq2}|F_k|=|F/F_1|.
\end{displaymath}
 Let $g=\max_{E\in F_1} w^{-}(E)$. Since $w^{-}(F_1)=\sum_{E\in F_1}w^{-}(E)\leq
 g|F_1|$, then
\begin{displaymath}
|F|/|F_1|\leq g+1.
\end{displaymath}
The weight function $w: K \rightarrow \mathcal {R}^{+}$ defined as
follows. For any arc $(E,E^{'})$ with $E^{'}\in F_k $, we know
$E^{'}$ is obtained by coalescing two cycles of $E$, and suppose the
length of these two cycles are $l_{1}$ and $l_2$, then define
$w(E,E^{'})=(l^{-1}_1+l^{-1}_2)f(k)$, where $f(k)$ is defined as in
Lemma 8. Then we have the following two claims.\\\\
\textbf{Claim 1}. \emph{For any $E\in F_k$, $k\geq 2$, $w^{+}(E)\geq
(2\alpha-1.5)n\beta f(k)\log n+2$.}\\\\
\textbf{Claim 2}. \emph{ For any $E\in F_k$, $k\geq 1$,
$w^{-}(E)\leq n\log
nf(k)$}.\\\\
By these two claims, set $\beta=1/(2\alpha-1.5)$. Then for $E\in
F_k$, $k\geq 2$, we have $w^{+}(E)-w^{-}(E)\geq 2\geq 1$ and
$g=\max_{E\in F_1} w^{-}(E)\leq n\log nf(1)\leq
(2\alpha-1.5)n^{1+1/(2\alpha-1.5)}$. Hence $|F|/|F_1|\leq g+1=
O(n^{1+1/(2\alpha-1.5)})$, which completes the proof.\ \ \
$\Box$\\\\
\textbf{Proof of Claim 1.} Let $E\in F_k$ be an 1-factor with $k$
cycles $\gamma_1, \cdots, \gamma_k$, of lengths $n_1, n_2, \cdots,
n_k$, $k\geq 2$. We proceed to bound $w^{+}(E)$. To show the lower
bound of $w^{+}(E)$, we need to count the number of arcs leaving
$E$. Suppose $(E,E^{'})$ to be such an arc. Let $\gamma=E\oplus
E^{'}$, $E^{'}\in F_{k-1}$, be the form $(x,x^{'},y,y^{'})$, where
$(x,x^{'}), (y,y^{'})\in E$ and $(y,x^{'}), (x,y^{'})\in E^{'}$.

First, we estimate the number of $\overline{C_4}$-type cycles
$\gamma$ for which $(x,x^{'})$ is contained in a particular cycle
$\gamma_i\in E$. We say that $\gamma$ is $rooted$ at $\gamma_i$.
Assume, for a moment, that the vertices $x, x^{'}$ is chosen. There
are at least $\alpha n-(n_i-1)$ ways to extend  the path first to
$y$ then to $y^{'}$ since the indegrees of  $x^{'}$ is at least
$\alpha n$. Denote $Y^{'}$ the set of all vertices $y^{'}$
reachable. Recall $N^{+}(x,G)$ is the set of neighbors $x$ points
to. Thus the number of ways of completing a $\overline{C_4}$-type
cycle $(x,x^{'},y,y^{'})$ is at least
\begin{displaymath}
\begin{split}
|N^{+}(x,G)| + |Y^{'}|-n &\leq \alpha n + (\alpha n-(n_i-1))  -n\\
&=2\alpha n-n_i-n+1.
\end{split}
\end{displaymath}
Hence the total number of $\overline{C_4}$-type cycles rooted at
$\gamma_i$ is at least $n_i(2\alpha n-n_i-n+1)$.

We are now poised to bound $w^{+}(E)$. Each arc $(E,E^{'})$ defined
by a $\overline{C_4}$-type $\gamma$ rooted at $\gamma_i$ has weight
at least $n_i^{-1}f(k-1)$, which, by Lemma 8, bounded below by
$(\beta\log n)(kn_i)^{-1}f(k)$, Thus\\
\begin{displaymath}
\begin{split}
w^{+}(E)&=\sum\limits_{E^{'}:(E,E^{'})\in K}w(E,E^{'})\\
&\geq \sum\limits_{i=1}^kn_i(2\alpha n-n_i-n+1)n_i^{-1}f(k-1)\\
&= \sum\limits_{i=1}^k(2\alpha n-n_i-n)f(k-1)+ k f(k-1) \\
&\geq (2\alpha kn-n-kn)(\beta \log n)k^{-1}f(k) + k f(k-1) \\
&=(2\alpha -1/k-1)(\beta \log n)f(k)n + k f(k-1)\\
&\geq (2\alpha -1.5)(\beta \log n)f(k)n + 2.
\end{split}
\end{displaymath}

For the first inequality, it seems we have overcounted the weight.
we explain the reason. When $(x, x^{'})$ is rooted at $\gamma_i$ and
$(y,y^{'})$ lies in some $\gamma_j$ if we extends $(x, x^{'})$ to
$(y,y^{'})$ to complete a $\overline{C_4}$-type cycle $\gamma=(x,
x^{'},y,y^{'})$, the contribution to the weight is only
$n_i^{-1}f(k-1)$ in the above inequality. Similarly, when $(x,
x^{'})$ is rooted at the same position as $(y,y^{'})$ in $\gamma_j$
and $(y,y^{'})$ lies in the same position as $(x, x^{'})$ in
$\gamma_i$, the contribution to the weight is $n_j^{-1}f(k-1)$. Plus
these two weight, $(n_j^{-1}+n_i^{-1})f(k-1)$ is exactly
$w(E,E^{'})$ needed to be considered by the definition of $w$, where
$E\oplus E^{'}=\gamma$. Hence, though each $\overline{C}_4$ cycle is
counted twice, the weight not. The last inequality follows
immediately from $k\geq 2$ and
$f(k-1)\geq 1$. \ \ \ $\Box$\\\\
\textbf{Proof of Claim 2.} For each $E\in F_k$, we now proceed to
bound $w^{-}(E)$. Let $(E^{'},E)$ be an arc in $K$. It is
straightforward to verify that the $\overline{C_4}$-type $\gamma=(x,
x^{'},y,y^{'})= E\oplus E^{'}$ must contain two edges $(x, x^{'})$
and $(y,y^{'})$ from a single $\gamma_i$ of $E$, and $(y,x^{'}),
(x,y^{'})\in E^{'}$. Removing these two edges from $\gamma_i$ leaves
a double of simple paths of lengths $p-1$ and $q-1$, where $p,q \geq
2$. For the case $p\neq q$ there are at most $n_i$ ways such that
$\gamma_{i}\oplus\gamma$ is a pair of cycles with length $p$ and
$q$, and $n_i/2$ ways such that $\gamma_{i}\oplus\gamma$ is a pair
of cycles with length $p$ and $q$ for the case $p=q$. Noting both
cases happen when $\gamma_i$ is contained in a complete sub digraph
of $G$ or $G$ is a complete digraph(Complete digraph is defined as
such a digraph that any two distinct vertices have edges pointing to
 each other). Hence\\
\begin{displaymath}
\begin{split}
w^{-}(E)&=\sum\limits_{E^{'}:(E^{'},E)\in K}w(E^{'},E)\\
&\leq \sum\limits_{i=1}^k n_i f(k)\sum\limits_{\substack{p>q\geq
2\\p+q=n_i}}(\frac{1}{p}+\frac{1}{q})+\frac{1}{2}\sum\limits_{i=1}^k
n_i f(k)\sum\limits_{\substack{p=q\geq
2\\p+q=n_i}}(\frac{1}{p}+\frac{1}{q})\\
&\leq \frac{1}{2}\sum\limits_{i=1}^k n_i
f(k)\sum\limits_{\substack{p,q\geq
2\\p+q=n_i}}(\frac{1}{p}+\frac{1}{q})\\
&=\frac{1}{2}\sum\limits_{i=1}^k n_i f(k)\sum\limits_{p=2
}^{n_i-2}(\frac{1}{p}+\frac{1}{n_i-p})\\
&=\sum\limits_{i=1}^k n_i f(k)\sum\limits_{p=2
}^{n_i-2}(\frac{1}{p})\\
&\leq\sum\limits_{i=1}^k n_i f(k)\log n_i\\
&\leq n \log n f(k). \ \ \ \Box
\end{split}
\end{displaymath}  \\\\\\
\textsl{4.2. Hardness of Counting Hamiltonian cycles in dense
digraphs}\\

We first declare the notation related to undirected graphs only
appears in this subsection and the notation related to digraph is
the same as that in the previous sections. Our reduction comes from
the undirected graph, hence notation for undirected graphs is
needed. Let $G$ be a simple $undirected$ graph with vertices
$\{1,2,\cdots,n\}$, where $n\geq 3$. The definition of a Hamiltonian
cycle of an undirected graph is a closed undirected path that visits
each vertex once and only once. We use the notation $m_1m_2\cdots
m_{n}m_1$ to denote a Hamiltonian cycle in an undirected graph
(recall $(m_{1},m_{2},\cdots,m_{n},m_{1})$ denotes a Hamiltonian
cycle in digraphs). The degree of a vertex in an undirected graph
$G$ is defined as the number of its neighbors. Let $\#$HC and
$\#$DHC be the problem of counting the number of Hamiltonian cycles
in undirected and directed graphs respectively. Now define a
symmetric digraph $G'$ corresponding to an undirected graph $G$ by
replacing each edge $(i,j)$ of $G$ with two directed edges $(i,j)$
and $(j,i)$. Let $H_G$ and $H_{G'}$ denote the set of the
Hamiltonian cycles in $G$ and $G'$ respectively. Let $\mathcal
{P}(H_{G'})$ denote the power set of $H_{G'}$. We will prove the
number of Hamiltonian cycles in an undirected graphs equals half of
the number of Hamiltonian cycles in its corresponding symmetric digraphs.\\\\
\textbf{Lemma 10.} (\cite{DFJ94}) \emph{$\#$HC is $\#$P-Complete,
even when restricted to graphs $G$ of minimun degree at least
$(1-\varepsilon)n$, for any
$\varepsilon>0$}\\\\
\textbf{Lemma 11.} \emph{Let $\mathcal{H}=m_1\cdots m_{n}m_1$ be a
Hamiltonian cycle in $H_G$. Then there are at least two Hamiltonian
cycles $(m_1,\cdots, m_{n}, m_1)$ and $(m_1, m_n,\cdots, m_1)$ in
$H_{G'}$. Define a map $\varphi$ from $H_G$ to $\mathcal
{P}(H_{G'})$ as follows:
\begin{displaymath}
\varphi(\mathcal{H})= \{(m_1, \cdots, m_{n}, m_1), (m_1, m_n,
\cdots, m_1)\}.
\end{displaymath}
Let $Im\varphi$ denote the image set of the map $\varphi$, and let
$\mathcal{H}'=m'_1\cdots m'_{n}m'_1$ be a different Hamiltonian
cycle from $\mathcal{H}$ in $H_G$. Then
\begin{displaymath}
\varphi(\mathcal{H}) \cap \varphi(\mathcal{H}')=\emptyset \qquad
\operatorname{and} \qquad \cup Im\varphi=H_{G'}.
\end{displaymath}}\\
\textbf{Proof.} Due to the symmetry of the digraph $G'$, and noting
$n\geq 3$, for any Hamiltonian cycle $(m_1,\cdots, m_{n}, m_1)$ in
$H_{G'}$, there must be a different Hamiltonian cycle $(m_1,
m_n,\cdots, m_1)$ in $H_{G'}$.  These two Hamiltonian cycles
obviously has a pre-imagine, the Hamiltonian cycle $m_1\cdots
m_{n}m_1$ in $H_{G}$. Note $(m_1,\cdots, m_{n}, m_1)$ is in
$\varphi(m_1\cdots m_{n}m_1)$. Hence, $\cup Im\varphi\supseteq
H_{G'}.$ Obviously, $\cup Im\varphi\subseteq H_{G'}$. Therefore
$$\cup Im\varphi=H_{G'}.$$ Suppose there are two different Hamiltonian
cycles $\mathcal{H}=m_1\cdots m_{n}m_1$ and $\mathcal{H}'=m'_1\cdots
m'_{n}m'_1$ in $H_{G}$.  Let $N_{\mathcal{H}}(m_i)$ denote two
neighbor vertices of vertex $m_i$ in $\mathcal{H}$. $\mathcal{H}$
and $\mathcal{H}'$ are different if and only if there exits a vertex
$\{m_i\}$=$\{m'_j\}$ such that $N_{\mathcal{H}}(m_i)\neq
N_{\mathcal{H}'}(m'_j)$. Hence $(m_1,\cdots, m_{n}, m_1)$ is
different from $(m'_1, \cdots, m'_{n}, m'_1)$ and $(m'_1, m'_n,
\cdots, m'_{2}, m'_1)$, that is $(m_1,\cdots, m_{n},
m_1)\notin\varphi(\mathcal{H}')$. Similarly, $(m_1, m_n, \cdots,
m_1)$ $\notin$ $\varphi(\mathcal{H}')$. Hence $\varphi(\mathcal{H})$
$\cap$
$\varphi(\mathcal{H}')$ $=\emptyset$. \ \ \  $\Box$\\\\
\textbf{Theorem 12.} \emph{$\#$DHC is $\#$P-Complete, even when the
digraph is $(1-\gamma)n$ dense,
$0<\gamma<.5$.}\\
\textbf{Proof.} Lemma 11 shows the number of Hamiltonian cycles in
an undirected graph is half of the number of Hamiltonian cycles in
its corresponding symmetric digraph. Hence by Lemma 10, $\#$DHC in
$(1-\gamma)n$ dense digraphs
is $\#$P-Complete, for any $0<\gamma<.5$.\ \ \ $\Box$\\\\
\textbf{\large{5. Conclusions and Discussions}}

The results in this paper show that for relatively dense digraphs,
approximating the number of Hamiltonian cycles or generating
weighted Hamiltonian cycles exactly from their correct distribution
can be accomplished in $O(n^{2.5+.5/(2\alpha-1)+2/{(4\alpha-3)}})$
time. This is an improvement in running time by a factor of
$n^{4.5}(\log n)^4$ for $.85n$ dense digraphs. Counting the number
of Hamiltonian cycles in such digraphs is shown to be
$\#$P-Complete.

Estimating the Hamilton of a 0-1 matrix to within a factor of
$1+\varepsilon$ with probability at least $1-\delta$, the running
time is
$$O(n^{2.5+.5/(2\alpha-1)+1/{(2\alpha-1.5)}}\varepsilon^{-2}\log(\delta^{-1})).$$

It is known \cite{BM76} that $0.5n$ dense digraphs contain
Hamiltonian cycles. Our algorithm presented in this paper is shown
to be an FPRAS for $0.75n$ dense problems. Hence a gap still
remains. We can extend the definition  $\overline{C_4}$ in the proof
of Theorem 9, as shown by Figure 1. Similarly that can also be done
to $\overline{C_6}$, $\overline{C_8}$. However it seems unlikely to
obtain any better bounds than that by $\overline{C_4}$ in this way.
This gap is  left open here.



%

\end{document}